\documentclass[%
 reprint,
superscriptaddress,
 amsmath,amssymb,
 aps,
 prb,
floatfix,
]{revtex4-2}

\usepackage{graphicx}
\usepackage{dcolumn}
\usepackage{multirow}
\usepackage{color}
\usepackage{hhline}
\usepackage{ulem}
\usepackage{bm}
\usepackage{physics}
\usepackage{hyperref}
\renewcommand{\r}[1]{\Vec{r}_{#1}}
\newcommand{\cbraket}[2]{\langle\langle#1|#2\rangle\rangle}
\newcommand{\pvec}[1]{\vec{#1}\mkern2mu\vphantom{#1}}

\begin{document}

\preprint{APS/123-QED}

\title{Pauli crystal melting in shaken optical traps}

\author{Jiabing Xiang}
\affiliation{Institute of Physics, Albert-Ludwig University of Freiburg, Hermann-Herder-Strasse 3, 79104 Freiburg, Germany}
\author{Paolo Molignini}%
\affiliation{ Cavendish Laboratory, Department of Physics, University of Cambridge, CB3 0HE Cambridge (United Kingdom)}%
\author{Miriam Büttner}
\affiliation{Institute of Physics, Albert-Ludwig University of Freiburg, Hermann-Herder-Strasse 3, 79104 Freiburg, Germany}
\author{Axel U. J. Lode}
\email{E-mail: auj.lode@gmail.com}
\affiliation{Institute of Physics, Albert-Ludwig University of Freiburg, Hermann-Herder-Strasse 3, 79104 Freiburg, Germany}

\date{\today}

\begin{abstract}
Pauli crystals are ordered geometric structures that emerge in trapped noninteracting fermionic systems due to their underlying Pauli repulsion. The deformation of Pauli crystals - often called \textit{melting} - has been recently observed in experiments, but the mechanism that leads to it remains unclear.
We address this question by studying the melting dynamics of $N=6$ fermions as a function of periodic driving and experimental imperfections in the trap (anisotropy and anharmonicity) by employing a combination of numerical simulations and Floquet theory.
Surprisingly, we reveal that the melting of Pauli crystals is not simply a direct consequence of an increase in system energy, but is instead related to the trap geometry and the population of the Floquet modes. We show that the melting is absent in traps without imperfections and triggered only by a sufficiently large shaking amplitude in traps with imperfections.

\end{abstract}

\maketitle

\section{\label{sec:introduction}Introduction}
Correlations between particles underlie many complex phenomena in quantum many-body systems, from the quantum Hall effect \cite{tsui1982,stormer1999,bolotin2009} to high-temperature superconductivity \cite{bednorz1986,anderson1987,dagotto1994}. 
The exploration of the type and origin of correlations is thus essential to understand the complexity of quantum many-body systems. 
Of particular interest to our research is the fact that correlations can emerge as self-ordering structures in the spatial distribution and lead to intriguing crystalline phases of matter.
More commonly however, solid crystalline phases are formed as a result of more conventional inter-particle interactions.
Already in 1934 Eugene Wigner theorized that a solid phase of electrons will form when the density of electrons is so low that their wave functions do not overlap due to long-range Coulomb interactions~\cite{wigner1934}, and his prediction was verified experimentally decades later~\cite{Grimes1979,Jiang1989}.
More recently, other intriguing quantum crystalline phases have been obtained experimentally, such as in dipolar quantum gases~\cite{lahaye2007,Kadau2016}, Rydberg atoms~\cite{Schauss2015,lechner2013}, and Bose-Einstein condensates in optical cavities~\cite{leonard2017}.
\par
Pauli crystals are yet another crystalline phase that was recently realized thanks to rapid developments in the preparation and control of ultracold atoms~\cite{Gajda2016,Rakshit2017,Batle2017,Holten2021}.
Pauli crystals are geometrically ordered structures that form at extremely low temperatures when noninteracting fermions are constrained by trapping potentials.
Whereas the emergence of other crystalline phases is typically driven by particularly engineered repulsive interactions, Pauli crystals rely solely on the more fundamental quantum statistics of fermions.
The competition between the Pauli exclusion principle and the attractive trap potential leads to N-body correlations in the relative positions of these fermions, and manifests itself in the specific self-ordering structures of Pauli crystals.
\par
The detection of Pauli crystals requires rigid conditions in experiments since the spatial geometric structure of correlations that relies on quantum statistics can be easily destroyed by perturbations such as trap imperfections.
The first detection of Pauli crystals was not realized until 2020~\cite{Holten2021}.
In this experiment, three or six $^6$Li atoms are trapped in their ground state in a quasi-two-dimensional harmonic trap.
The detection of the crystalline structure is performed by the simultaneous measurement of all atom positions (single-shot imaging) via time-of-flight (TOF) expansion techniques \cite{murthy2014} and single atom detection fluorescence imaging schemes \cite{bergschneider2018}. 
After image processing of many ($\sim 3000$) single-shot images, the crystalline structures can be observed.
\par
Heating effects are particularly detrimental to the stability of Pauli crystals. 
When the system heats up, the spatial extension of the particles increases.
The contribution from the N-body correlations in the excited states blends with the ones in the ground state and causes deformations in the crystal structure, a phenomenon often termed \textit{melting}~\cite{Rakshit2017}.
In the experiment, to investigate the effect of heating on Pauli crystals, the six-fermion system is subject to a sinusoidal modulation of the trap frequency at twice the original trap frequency with variable amplitudes. 
The measurements verified the melting phenomenon: with an increase in the system energy, the geometric structure becomes progressively more deformed until the crystalline phase is lost completely. \par
The dynamical process of Pauli crystal melting raises some intriguing questions.
First of all, the precise role of the trap imperfections on the melting process is still unknown. 
Deviations from a perfect harmonic trap can influence the quantum dynamics and will be amplified during a long time modulation. 
They must be therefore taken into account to study the stability of Pauli crystals.
Furthermore, it is a priori not clear whether Pauli crystal deformations are due purely to a generalized increase of the system energy or if driving induces more complex many-body excitations.
\par
In this work, we address both questions by unveiling how the geometric structures of Pauli crystals change in different dynamical potentials: an isotropic harmonic trap $V_{\text{h}}$, an anisotropic harmonic trap $V_{\text{ani}}$, an isotropic anharmonic trap $V_{\text{anh}}$, and a Gaussian trap $V_{\text{Gau}}$ that combines both types of imperfections (anisotropy and anharmonicity).
We discover that the occurrence of melting depends heavily on the trap imperfections. 
For the isotropic harmonic trap, we observe a melting-resistant behavior, where the particular geometric structure is still visible even when the system energy increases to ten times the energy measured in the experiment.
For the other traps with imperfections, the melting phenomenon can be reproduced with the same modulation given in the experiment.
To understand the absence of melting in the isotropic harmonic trap, we compare the wave functions of the melting and non-melting systems via the excitation spectra.
We find that the superposition of different quasienergy states can lead to a strong reduction of Pauli correlations in the time evolution.
For the isotropic harmonic trap, the wave function is significantly dominated by the ground state even when modulations are present, and the system maintains the ground-state correlation in the dynamical process.
For the other traps, instead, the population of other Floquet modes progressively destroys the Pauli crystal structure.
Our work contextualizes Pauli crystal melting in terms of geometric and dynamical factors and should serve as a blueprint on how to better stabilize this exotic phase of matter in future studies.
\par
The rest of this paper is organized as follows. 
In Sec.~\ref{sec:systems}, we present the system that we consider and its Hamiltonian description with the four different trapping potentials.
The definition of the configuration density and of the recognition function used to quantify the deformation extent is presented in Sec.~\ref{subsec:con_den} and Sec.~\ref{subsec:R}, separately. 
Thereafter, we provide a review of the numerical method that we employ for our simulations (MCTDH-X~\cite{lode2019,lin2020,lode2016,fasshauer2016}) in Sec.~\ref{subsec:mctdhx} and the description of the autocorrelation function via Floquet theory in Sec.~\ref{subsec:floquet}. 
The simulation results and melting mechanism analysis are discussed in Sec.~\ref{sec:results}. 
First, we distinguish between melting and non-melting systems with the configuration density and the recognition function. 
Then, the relationship between the system energy and the melting in the dynamical process is analyzed. 
Finally, the melting mechanism is revealed by comparing the Fourier transform of the autocorrelation function in the melting systems with the non-melting ones. 
Sec.~\ref{sec:conclusions} summarizes our findings and presents an outlook on possible future studies.
\section{\label{sec:systems}System}
We consider $N=6$ fermions in an optical trap, described by the Hamiltonian
\begin{equation}
    H=\sum_{i=1}^N \left[ -\frac{1}{2}\nabla^2_i+V(\pvec{r}_i,t) \right],
\end{equation}
where $\pvec{r}_i=(x_i,y_i)$ is the coordinate of particle $i$.
Throughout this paper we use dimensionless units, for which energy and distance are measured in units of $\hbar\omega_0$ and $\sqrt{\hbar/(m\omega_0)}$, respectively. 
The initial radial frequency $\omega_0$ is set to $2\pi\times983(5)$ Hz, as in the experiment\cite{Holten2021}.
\par
The simplest generic trap that we consider is an isotropic harmonic potential with frequency $\omega_x=\omega_y$,
\begin{equation}
    V_\mathrm{h}(\r,t)=\frac{1}{2}\omega_r^2(t)\pvec{r}^2, \label{Visoh}
\end{equation}
which is the idealized model for the experiment.
The $n$-th energy level of a fermion making up the $n$-th energy shell in the two-dimensional isotropic harmonic trap is $(n+1)$-fold degenerate.
\par

The fermionic system is initially prepared in the ground state, where the geometric structure of Pauli crystals can be observed distinctly. 
In order to melt the Pauli crystal, much like in the experiment~\cite{Holten2021}, we then introduce a time-periodic modulation of the confining trap frequency in the radial direction
\begin{equation}
    \omega_r(t)=\omega_0+A\sin(\omega_t t),
    \label{modulation}
\end{equation}
where $A$ is the shaking amplitude, 
$\omega_0$ is the initial trap frequency, and $\omega_t$ is the modulation frequency.
We set the modulation frequency $\omega_t=2\omega_0$, since
this is exactly the transition frequency from any energy shell to two shells up.

In the experiment, there are both anisotropy and anharmonicity corrections to the trap due to the small size of the optical tweezer and a finite coupling with angular momentum in $z$-direction~\cite{Holten2021}. 
To study the effect of trap imperfections on the system dynamics, we first study these corrections separately.
The anisotropic trap is given by
\begin{equation}
    V_{\text{ani}}(\r,t)=\frac{1}{2}\omega_r^2(t)\left(\gamma x^2+\frac{y^2}{\gamma}\right),\label{Vani}
\end{equation}
with the anisotropy parameter $0<\gamma<1$.
The anharmonic trap is instead described by
\begin{equation}
    V_{\text{anh}}(\r,t)=\frac{1}{2}\omega_r^2(t)\pvec{r}^2+B\pvec{r}^4,\label{Vanh}
\end{equation}
with the anharmonicity parameter $B>0$. 
All kinds of corrections to the trap split the degenerate energy levels of the isotropic harmonic oscillator.
As a result, the complexity of the energy level structures in the imperfect traps will produce different system dynamics in the modulation process. \par

Finally, we analyze a Gaussian trap that combines both types of corrections~\cite{bayha2020}:
\begin{equation}
    V_{\text{Gau}}(\r,t)=\frac{B^\prime}{2}\omega^\prime_r(t)\times\left[1-e^{\left(\gamma^\prime x^2+\frac{y^2}{\gamma^\prime}\right)/B^\prime}\right].\label{Vgau}
\end{equation}
For the simulation of the Gaussian trap, we use the parameter values measured in the experiment: $B^{\prime} =20$ and $\gamma^\prime=0.99$.

\begin{figure}[h]
    \centering
    \includegraphics[scale=0.91]{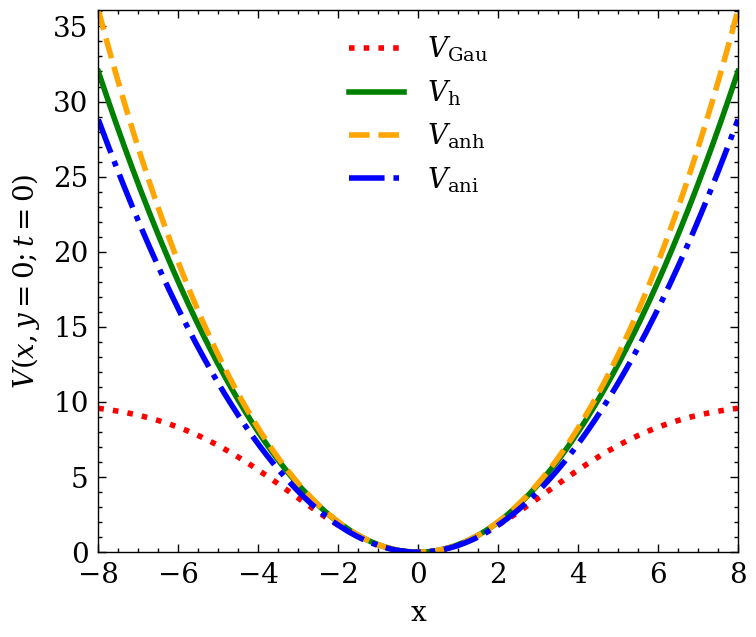}
    \caption{Comparison of the trapping potentials investigated in this work. The dotted red line indicates the Gaussian trap Eq.~\eqref{Vgau} with $B'=20$ and  $\gamma'=0.99$. The continuous green line corresponds to the isotropic harmonic trap $V_h$ Eq.~\eqref{Visoh}. The dashed yellow line marks the isotropic anharmonic trap $V_{\text{anh}}$ Eq.~\eqref{Vanh} with an anharmonicity $B=0.001$. 
    The dash-dotted blue line represents the anisotropic harmonic trap $V_{\text{ani}}$ Eq.~\eqref{Vani} with anisotropy  $\gamma=0.90$.}
    \label{system:traps}
\end{figure}

As shown in Fig.~\ref{system:traps}, there is an evident discrepancy between the Gaussian potential and the other potentials away from the bottom of the trap.
This leads to a distinct difference between the energy levels and the transition frequency of the Gaussian trap and those of the isotropic harmonic trap.
This is consistent with a trial computation that demonstrates that the modulation of $2\omega_0$ can not excite the fermionic system efficiently.
Since the fermionic system in the Gaussian trap is out of resonance with the modulation frequency of $2\omega_0$,
we adopt the transition frequency from the lowest energy shell to two shells up as the modulation frequency, which we calculated numerically to be $\omega_t=(E_3-E_0)/\hbar=1.843\omega_0$ (cf. Table~\ref{system:en-gau}).
%
\begin{table}[h]
    \centering
    \begin{tabular}{|c|c|}
    \hline
    Energy shell number & Energy after splitting/$\hbar\omega_0$ \\
    \hline
    1 & $E_0=0.975$\\
    \hline
    2 & $E_1=1.919$, $E_2=1.928$\\
    \hline
    3 & $E_3=2.818$, $E_4=2.846$, $E_5=2.852$\\
    \hline
    \end{tabular}
    \caption{Energy levels for the lowest three shells in the Gaussian trap calculated with MCTDH-X. 
    Due to the anisotropy and anharmonicity of the Gaussian trap, the degeneracy of the energy shells present in the isotropic harmonic trap is split.}
    \label{system:en-gau}
\end{table}

\section{\label{sec:methods}Methods}
After introducing the system hosting the Pauli crystals, we present the analytical and numerical methods used to study their dynamics and melting.

\subsection{Configuration density}\label{subsec:con_den}
Despite the system setup enabling the Pauli principle to cause the particles to form geometric structures, these are concealed when examining  single-shot images. 
This is due to the rotational symmetry of the problem, which randomizes the particle position and configuration orientation in each single-shot image.
To reveal the correlation effects of Pauli crystals we then need to perform an additional image processing step as illustrated in Ref.~\cite{Gajda2016}.
We record single-shot images in such a way that only the information of the relative positions remains, while other details such as the position of the center of mass and the orientation of the configuration in space are discarded. 
The ensemble of post-processed single-shot images is called \textit{configuration density} $\mathcal{S}_{N_s}$ (see Appendix~\ref{app::image}). 

\begin{figure}[ht]
    \includegraphics[scale=0.29]{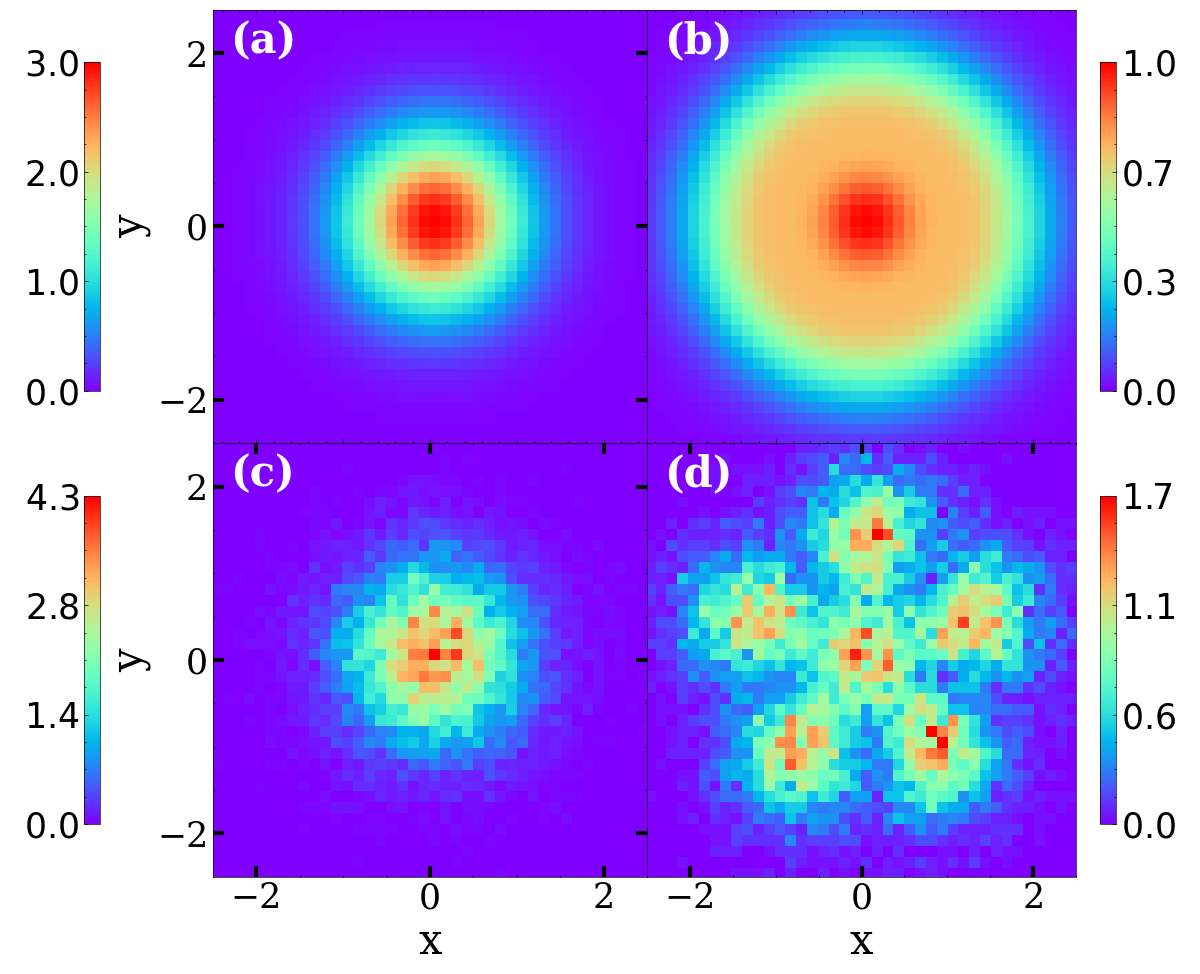}
    \caption{The one-body and configuration density of six fermions and bosons in the ground state of an isotropic harmonic trap. The unit of the colorbar is $10^{-2}$. \textbf{(a)} One-body density of the six-boson system. \textbf{(b)} One-body density of the six-fermion system. \textbf{(c)} Configuration density of the six-boson system. \textbf{(d)} Configuration density of the six-fermion system.}
    \label{method:con-den}
\end{figure}

To illustrate this procedure, Fig.~\ref{method:con-den} compares the one-body density to the configuration density for a  six-boson and a six-fermion noninteracting system in the ground state of an isotropic harmonic trap. 
In the one-body density of the six-boson system (Figs.~\ref{method:con-den}(a,b)), since the particles are noninteracting, all the bosons are localized at the minimum of the potential trap. 
In contrast, the Pauli exclusion principle leads to fermions effectively repelling each other. 
Hence, the one-body density of the six-fermion system is more extended in space than its bosonic counterpart. 
While the one-body density of the fermionic system can reveal the most probable position of a single particle, it does not reveal information about statistical correlations between particles, which is contained in the $N$-body density.
The $N$-body density is a higher-dimensional quantity with obvious visualization constraints.
However, as shown in Figs.~\ref{method:con-den}(c,d), its information is accessible in the configuration density.
As the non-interacting bosons behave as a single entity and lack correlations, the configuration density does not display any additional structure. 
The configuration density of the six-fermion system, on the other hand, reveals the correlation effects.
The fermions arrange themselves into a configuration that minimizes the states' overall energy within the interplay between Pauli repulsion and trap confinement, where one particle is localized in the center and the others form a regular pentagon in an outer shell.
In other words, the configuration density offers a way of extracting and visualizing the information contained in the $N$-body density in a practical way.

\par 

\subsection{Recognition function}\label{subsec:R}
When a Pauli crystal melts, the particular geometric structure of a configuration density like the one shown in Fig.~\ref{method:con-den}(d) becomes blurred and can even vanish.
The deformation of the geometric structure indicates a breakdown of the correlation between fermions at different positions.
To quantify the blurring of the geometric structure and the amount of melting, in this section we propose a \textit{recognition function} $R$ that can be derived from the one-body density and the configuration density.\par

By comparing melting and non-melting systems, or alternatively the ground-state of bosonic and fermionic noninteracting systems (see Fig.~\ref{method:con-den}), we notice that the geometric structure of the configuration density is almost isotropic and close to the corresponding one-body density when the effect of the Pauli exclusion principle is weak or absent~\footnote{Note that the slight fluctuations in the angular distribution are the result of the discretization error in the coordinate grid.}.
In contrast, the angular distribution of the configuration density is highly non-uniform for the fermionic ground state in the Pauli crystal phase. 
\begin{figure}[h]
    \centering
    \includegraphics[scale=0.6]{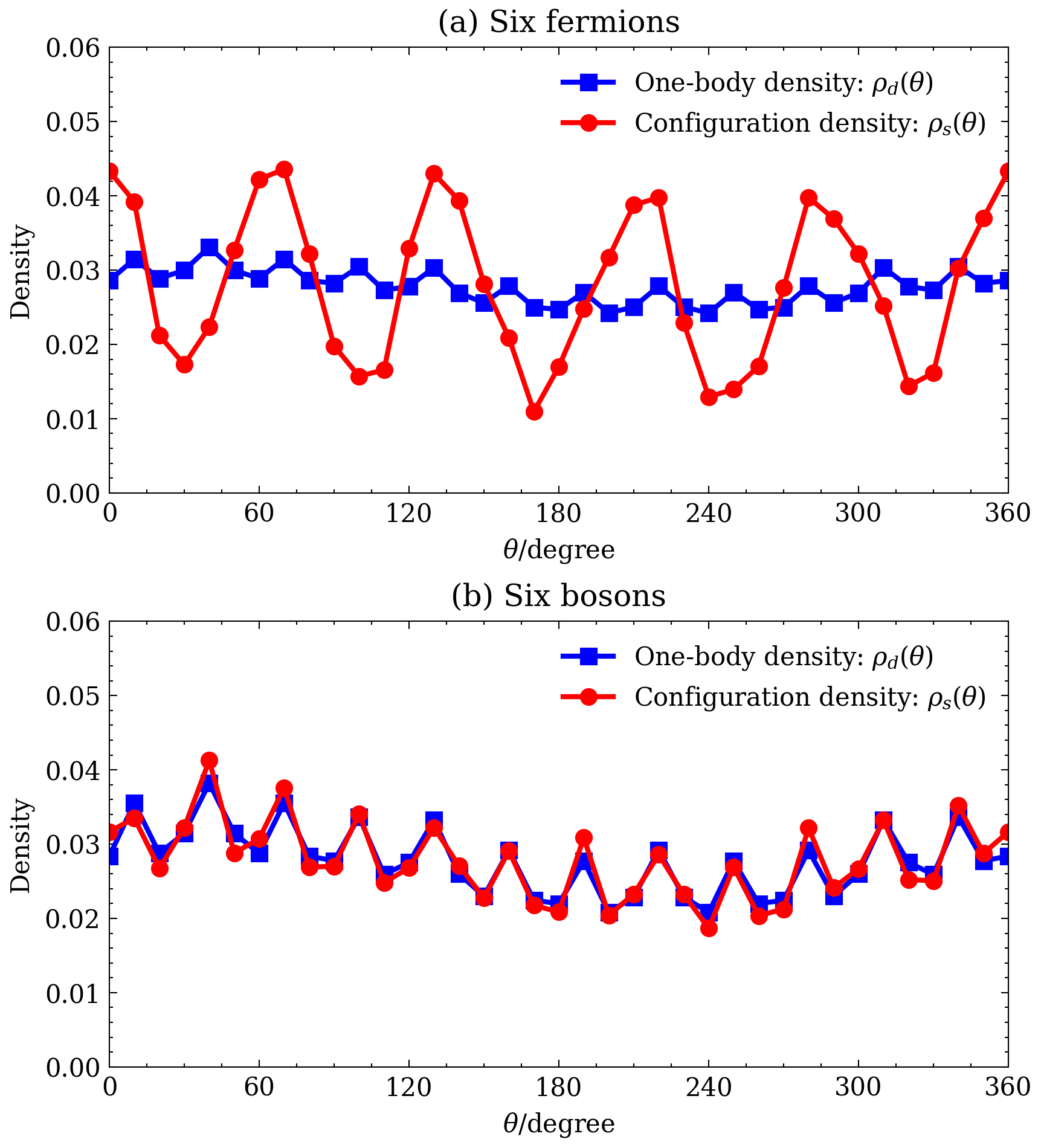}
    \caption{Pattern recognition for the six-particle systems. \textbf{(a)} For the ground-state fermion system, the angular distribution of the configuration density is different from its one-body density distinctly. \textbf{(b)} For the ground-state boson system, the angular distribution of the configuration and one-body densities is almost the same.}
    \label{method:ang-dis}
\end{figure}
Hence, we can use the difference between the angular distributions of the configuration density $\rho_s(\theta)=\int_0^{r_\text{max}} \:  \mathcal{S}_{N_s}(\Vec{r})r \mathrm{d}r $ and of the one-body density $\rho_d(\theta)=\int_0^{r_\text{max}} \: \rho(\Vec{r})r \mathrm{d}r $ to distinguish the two systems (see Fig.~\ref{method:ang-dis}). 
This difference can be described by the variance of each angular distribution $\mathbf{Var}(\rho_{s/d})=\int_0^{2\pi} (\theta-\Bar{\theta}_{s/d})^2\rho_{s/d} \mathrm{d}\theta$, where
$\Bar{\theta}_{s/d}=\int_0^{2\pi} \rho_{s/d}\ \mathrm{d}\theta$.
We then define the recognition function $R$ as the ratio of the variances between the two types of densities:
\begin{equation}
    R=\frac{\mathbf{Var}(\rho_s)}{\mathbf{Var}(\rho_d)}.
\end{equation}
Table~\ref{method:rec-funs} illustrates the variances and the recognition functions for both the six-fermion and six-boson system. 
The recognition function of the uncorrelated six-boson system is close to 1.
In contrast, the recognition function for the six-fermion system is much larger, revealing the underlying correlation structure of the Pauli crystal.

We can use the recognition function to define a precise \textit{melting criterion}:
when the recognition function of a given fermionic system reduces to a value comparable to the one for a corresponding bosonic system, the crystal has melted and the correlation of the fermionic system is completely destroyed.

%
\begin{table}[h]
    \centering
    \begin{tabular}{|c|c|c|c|}
    \hline
    System & $\mathbf{Var}(\rho_d)$  & $\mathbf{Var}(\rho_s)$ & $R$ \\
    \hline
    Six-fermion & $5.12\times10^{-6}$ & $1.04\times10^{-4}$ & $20.24$\\
    \hline
    Six-boson & $2.01\times10^{-5}$ & $2.66\times10^{-5}$ & $1.32$\\
    \hline
    \end{tabular}
    \caption{The recognition functions $R$ for the six-particle systems.}
    \label{method:rec-funs}
\end{table}

\subsection{MCTDH method}\label{subsec:mctdhx}
To study the many-body ground state $\left|\Psi\right>$ and its time evolution in the noninteracting fermionic system in different traps, we solve the many-body time-dependent Schr\"odinger equation,
\begin{equation}
    \hat{H}\ket{\Psi}=i\partial_t \ket{\Psi},
\end{equation}
using the MultiConfiguration Time-Dependent Hartree approach for indistinguishable particles (MCTDH method), implemented in the software MCTDH-X~\cite{lode2019,lin2020,lode2016,fasshauer2016}.
In MCTDH-X, the many-body wave function is expanded as a  linear combination of many-body states,
\begin{align}
    \ket{\Psi}&=\sum_{\Vec{n}}C_{\Vec{n}}(t)\ket{\Vec{n};t},\\
    \ket{\Vec{n};t}&=\frac{1}{\sqrt{N!}}\prod_{i=1}^M[\hat{b}^\dagger_i(t)]^{n_i}\ket{\text{vac}},
\end{align}
where both the coefficients $C_{\Vec{n}}(t)$ and the  many-body states $\ket{\Vec{n};t}$ are time dependent.
The many-body basis is composed of a total number of $M$ single-particle orbitals, $\Phi_i(\vec{r},t)=\bra{\vec{r}}\hat{b}^\dagger_i(t)\ket{\text{vac}}$.
MCTDH-X employs the time-dependent variational principle~\cite{mclachlan1964} to optimize both the coefficients $\{C_{\Vec{n}}(t)\}$ and the orbitals $\{\Phi_i(\vec{r},t)\}$
and obtain the numerically accurate solution for the many-body wave function of the time-dependent system.
In our computations, we employ $M=7$ orbitals for the many-body basis of the $N=6$ fermionic system, which allows us to separately describe each fermionic single-particle state.

For the wave function at each time $t$, we generate 3'000 single-shot images~\cite{sakmann2016,lode2017}.
A single-shot image is a simultaneous imaging measurement of the positions of all particles in a quantum state.
In MCTDH-X simulations, the many-body wave function $\ket{\Psi(\r{1},\cdots,\r{N});t}$ is computed at every time step.
The single-shot images are produced by drawing random samples from the $N$-body probability density $P(\r{1},\cdots,\r{N};t)=\abs{\Psi(\r{1},\cdots,\r{N};t)}^2$.
We then use the single-shot images to produce the configuration densities in the current quantum state and calculate derived quantities such as the recognition function.
Note that the one-body density and the autocorrelation function (see next section) are instead calculated directly from the many-body wave function (e.g. by Eq.~\eqref{App:one-body}).

\subsection{Autocorrelation function and Floquet theory}\label{subsec:floquet}
Floquet theory \cite{Shirley1965,Manakov1986,Sambe1973,Grossmann1991,Chu2004} is now a staple in the study of system dynamics driven by periodically time-dependent fields.
By means of Floquet theory, 
we can expand the wave function of the dynamical system as a superposition of simpler quasienergy states.
By analyzing the quasienergy state compositions of the many-body wave functions, we can better investigate the reason for the different dynamical behaviors in different contexts.
\par

A very useful quantity that can be expanded by means of Floquet theory and that gives insights into the nature of the many-body state is the autocorrelation function.
The autocorrelation function measures the overlap between the time-evolved wave function $\ket{\Psi(t)}$ and its initial wave function $\ket{\Psi(0)}$:
\begin{equation}
    a(t)=\bra{\Psi(0)}\ket{\Psi(t)}.
\end{equation}
From the Fourier transform of the autocorrelation function we can extract the energies and the relative population of the states involved in the system dynamics. \par

Generally, calculating the autocorrelation function of a time-dependent system is analytically intractable. 
However, for a system with a time-periodic Hamiltonian $\hat{H}_\tau(\vec{r},t)=\hat{H}_0(\vec{r})+\hat{V}_\tau(\vec{r},t)$ with $\hat{V}_\tau(\vec{r},t+\tau)=\hat{V}_\tau(\vec{r},t)$ and period $\tau$ ($\omega=2\pi/\tau$), we can obtain some analytical results by employing Floquet theory \cite{Shirley1965,Manakov1986,Sambe1973,Grossmann1991,Chu2004}.

\par
With the expression of the wave function derived from the Floquet theorem (see Eq.~\eqref{App:floquet-expansion} in the Appendix), the autocorrelation function can be written as
\begin{equation}
    a(t)=\sum_{j^\prime,j}\sum_{n^\prime,n}W_{j j^\prime n n^\prime} e^{-i(\epsilon_{\alpha(j)}+n\omega) t},
\end{equation}
where
\begin{equation}
    W_{j j^\prime n n^\prime}=A_{j^\prime}^*A_j\int \mathrm{d} \vec{r} \: {\varphi_{\alpha(j^\prime)}^{n^\prime\ *}}(\vec{r})\varphi_{\alpha(j)}^{n}(\vec{r}).
\end{equation}
The Fourier transform of $a(t)$, the so-called excitation spectrum, correspondingly reads
\begin{equation}
    \tilde{a}(\Omega)=\sum_{j^\prime,j}\sum_{n^\prime,n} W_{j j^\prime n n^\prime} \delta(\Omega-(\epsilon_{\alpha(j)}+n\omega)).
\end{equation}
The excitation spectrum is a sum of Dirac delta functions centered at the quasienergies $\epsilon_{\alpha_j}$.
Hence, the system dynamics under the influence of a time-periodic Hamiltonian $\hat{H}(t)$ can be regarded as the evolution of a superposition of quasienergy states that represent peaks in the Fourier transform of the autocorrelation function. 
We can thus understand the origins of the melting mechanism by analyzing the composition of the wave function through the excitation spectrum.

\section{\label{sec:results}Results}
\begin{figure*}[ht]
    \centering
    \includegraphics[width=\textwidth]{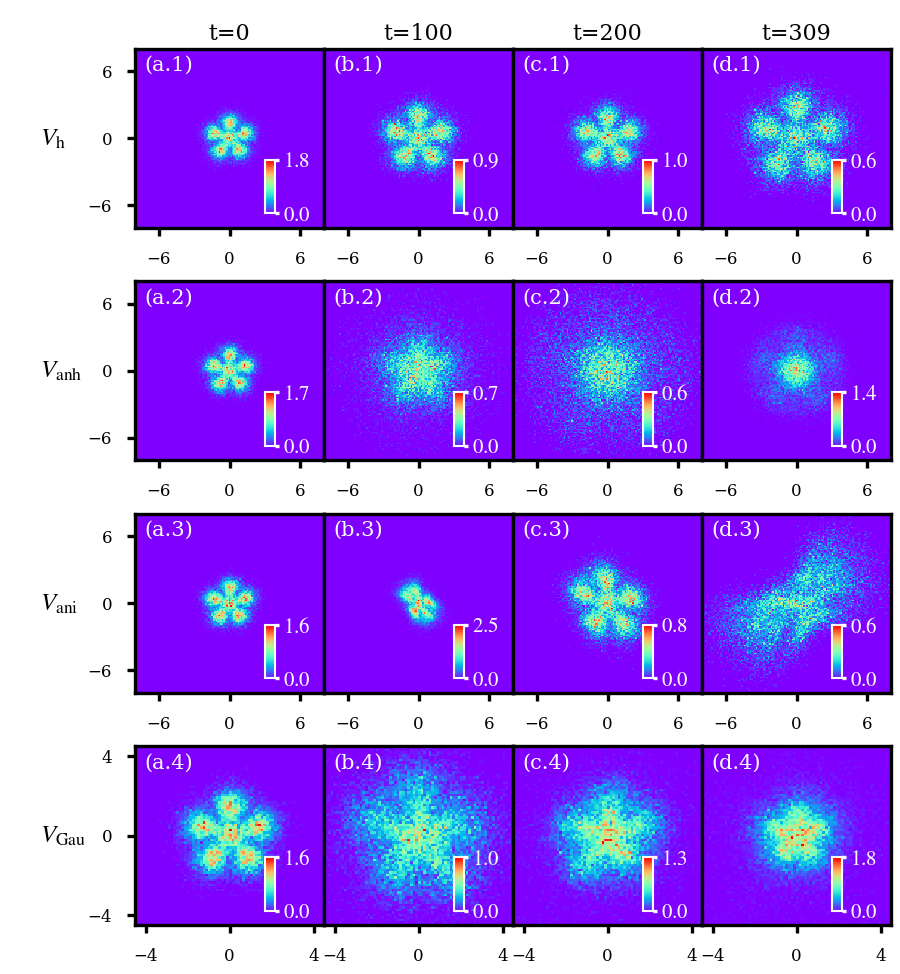}
    \caption{Configuration densities for the six-fermion system in the four traps. Panels \textbf{(a)-(d)} display the configuration density at the initial time $t=0$ (ground state), $t=100$, $t=200$ and $t=309$ (end of the modulation). The unit of the colorbar is $10^{-2}$. \textbf{(a.1)-(d.1)} Isotropic harmonic trap with shaking amplitude $A=0.01$. \textbf{(a.2)-(d.2)} Isotropic anharmonic trap with shaking amplitude $A=0.1$ and anharmonicity $B=0.001$. \textbf{(a.3)-(d.3)} Anisotropic harmonic trap with shaking amplitude $A=0.09$ and anisotropy $\gamma=0.90$. \textbf{(a.4)-(d.4)} Gaussian trap with shaking amplitude $A=0.25$.}
    \label{result:con_dens}
\end{figure*}
\begin{figure*}[th]
    \centering
    \includegraphics[width=0.95\textwidth]{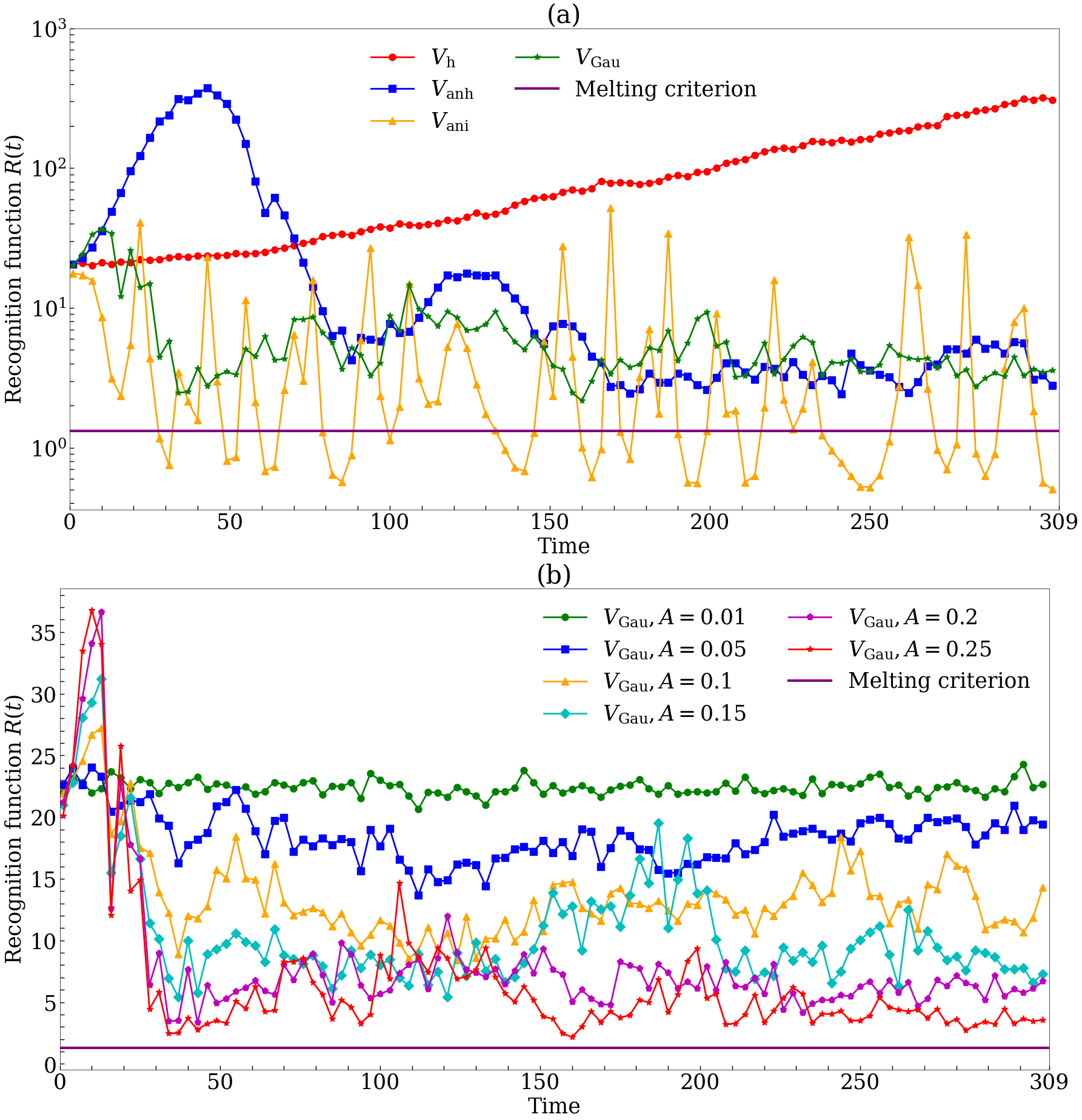}
    \caption{The time evolution of recognition functions $R(t)$ for the six-fermion system in the four traps. \textbf{(a)} Recognition function $R(t)$ averaged over every third period for the six-fermion system in each of the four traps considered in this work. The shaking amplitude $A$ for the six-fermion system in the isotropic harmonic trap $V_\text{h}$, isotropic anharmonic trap $V_\text{anh}$, anisotropic harmonic trap $V_\text{ani}$ and Gaussian trap $V_\text{Gau}$ are $0.01,0.1,0.09$, and $0.25$, respectively. \textbf{(b)} Recognition functions $R(t)$ for the six-fermion system in the Gaussian trap $V_{\text{Gau}}$ with different shaking amplitudes $A$.}
    \label{result:rec}
\end{figure*}

\begin{figure*}[th]
    \centering
    \includegraphics[width=0.95\textwidth]{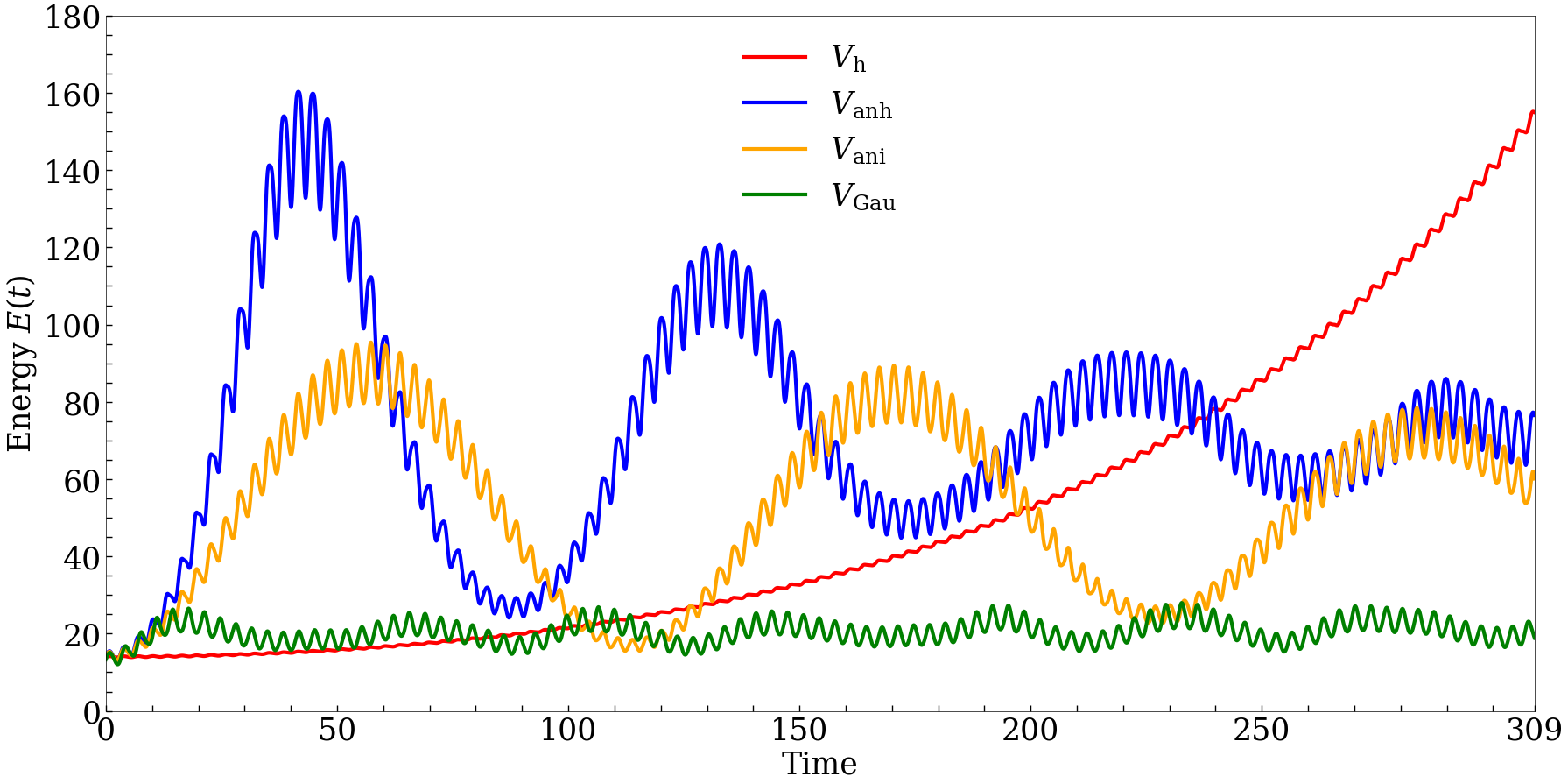}
    \caption{The time evolution of recognition functions $R(t)$ and system energies $E(t)$ for the six-fermion system in the four traps. The shaking amplitude $A$ for the six-fermion system in the isotropic harmonic trap $V_\text{h}$, isotropic anharmonic trap $V_\text{anh}$, anisotropic harmonic trap $V_\text{ani}$ and Gaussian trap $V_\text{Gau}$ are $0.01,0.1,0.09$, and $0.25$, respectively.}
    \label{result:Energy}
\end{figure*}

We now address the dynamics of Pauli crystal melting with the methods presented in the previous sections.

\subsection{Melting and non-melting system}
\begin{table}[th]
    \centering
    \begin{tabular}{|c|c|c|c|}
    \hline
    Trap & Anisotropy & Anharmonicity & Melting  \\
    \hline
    $V_{\text{h}}$ & No & No & No\\
    \hline
    $V_{\text{anh}}$ & No & Yes & Yes\\
    \hline
    $V_{\text{ani}}$ & Yes & No & Yes\\
    \hline
    $V_{\text{Gau}}$ & Yes & Yes & Yes\\
    \hline
    \end{tabular}
    \caption{Melting and non-melting systems with six non-interacting fermions.}
    \label{tab3}
\end{table}
The overall computation process with MCTDH-X can be divided into three steps.
Firstly, we relax the system in the trap to the ground state.
Secondly, we propagate the wave function in the driven trap from $t=0$ to $t=309$, which corresponds to a $50\mathrm{ms}$ modulation time in the experiment.
Thirdly, we calculate from the many-body wave function at every time step the desired observables, such as system energy, one-body density, single-shot images and autocorrelation function.\par

We simulate the modulation process in the four different traps for different sets of parameters (see Appendix~\ref{app::units}).
By calculating the configuration density and recognition function, we can categorize the different setups as melting or non-melting, as summarized in Table~\ref{tab3}.
Within the time scales given, we observe melting in all the traps except for the isotropic harmonic trap, which implies that trap imperfections are a key reason for the melting process.\par

For each trap, Fig.~\ref{result:con_dens} presents a comparison of the configuration density of the six-fermion system at the initial time $t=0$ (ground state), $t=100$, $t=200$ and the end of the modulation $t=309$. 
For the isotropic harmonic trap $V_{\text{h}}$ shown in Figs.~\ref{result:con_dens}(a.1)-(d.1), we notice that the geometric structure of the pattern persists until the end of the modulation. 
Although the pattern is sometimes diluted as the configuration density expands and contracts, the pentagon structure can be identified distinctly. 
Therefore, modulation in the isotropic harmonic trap does not lead to appreciable Pauli crystal melting.
\par
In contrast, for the systems with imperfect traps, the configuration density behaves very differently.
Both in the isotropic anharmonic trap shown in Figs.~\ref{result:con_dens}(a.2)-(d.2) and in the anisotropic harmonic trap shown in Figs.~\ref{result:con_dens}(a.3)-(d.3), the initial ground state configuration density becomes completely blurred or distorted by the end of the modulation process.
For the case of the Gaussian trap shown in Figs.~\ref{result:con_dens}(a.4)-(d.4), the particular geometric structure of the configuration density also becomes completely destroyed, although the required shaking amplitude is larger than the other two imperfect traps.

The melting process becomes apparent also in the recognition function, depicted in Fig.~\ref{result:rec}(a).
Note that to show the trend of the recognition function with time and remove some of the naturally occurring oscillations, we only plot its mean value averaged over three consecutive time points in an oscillation period.
Only for the isotropic harmonic trap, the recognition function stays well above the melting criterion throughout the time evolution.
In the traps with imperfections, instead, the recognition functions eventually decrease, coming close or even below the melting criterion, which indicates the complete deformation of the Pauli crystal.
We also compare the recognition function in the Gaussian trap with different shaking amplitudes in Fig.~\ref{result:rec}(b).
As the shaking amplitude increases, the recognition function reduces to closer to the melting criterion.
We remark that, compared to the other traps, the modulation frequency $\omega^\prime_t$ in the Gaussian trap is less efficient in exciting the fermionic system.
This is illustrated also in Fig.~\ref{result:Energy}, where this inefficient excitation results in a much lower system energy $E(t)$.
Hence, a larger shaking amplitude is required for the Gaussian trap to involve more excited states in the system dynamics and melt the Pauli crystal, which is in agreement with experimental observations.

\begin{figure}[h]
    \includegraphics[scale=0.7]{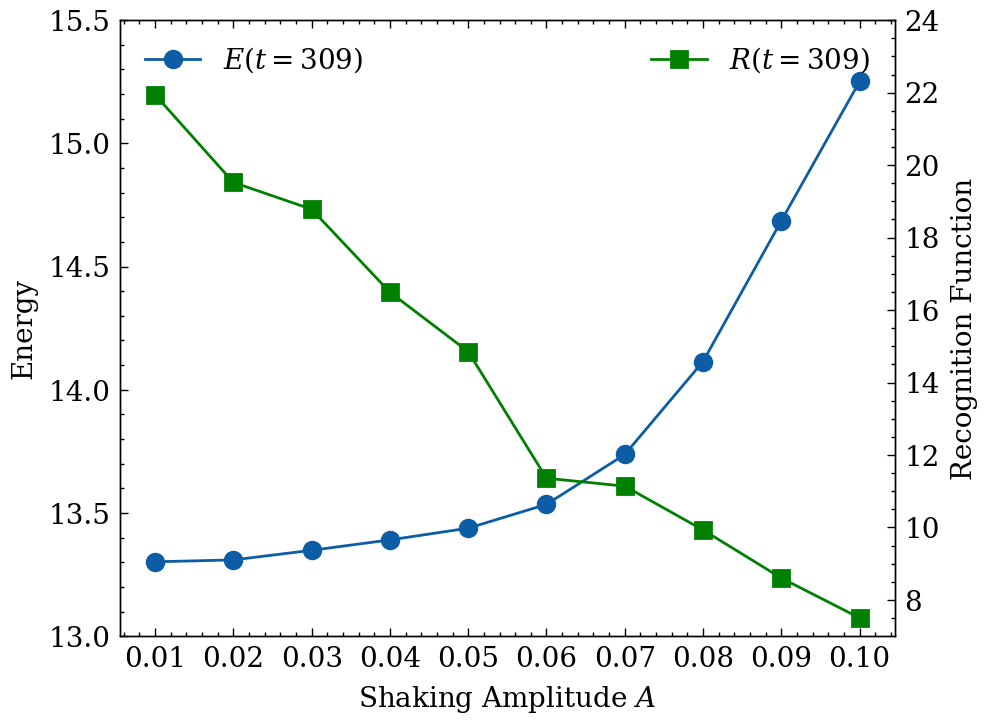}
    \caption{Melting of Pauli crystal in the Gaussian trap as evinced by the system energy and the recognition function at the end of modulation $t=309$ with different shaking amplitudes. When the shaking amplitude becomes larger, the system energy also increases, while the recognition function reduces.}
    \label{result:EwithR}
\end{figure}

\subsection{Energetics of the melting process}
We now present results about the accumulated energy $E=\bra{\Psi}\hat{H}\ket{\Psi}$ in each type of system and relate them to the dynamics of the Pauli crystal.
Intuitively, Pauli crystals should always melt when the system accumulates enough energy.
The simulation of the Gaussian trap, which by design is closest to the real experimental setup, reproduces the experimental observation that the particular geometric structure of Pauli crystals becomes more deformed as the system energy increases (compare Fig.~\ref{result:EwithR}).
However, our simulation results show that the transient behavior is more complex than this simplified picture.

Fig.~\ref{result:Energy} shows the system energies for the six-fermion system in the four traps.
Firstly, we can notice that the system energy $E(t)$ in the non-melting system (isotropic harmonic trap $V_\text{h}$) is larger than in any of the melting systems (the other traps) at the end of the modulation.
This indicates that the melting of Pauli crystals is more sensitive to the type of trap rather than to the system energy.
Another indication that the total energy is not a good indicator for the melting comes from analyzing the system energy and the recognition function for the six fermions in the isotropic anharmonic trap (blue lines in Fig.~\ref{result:rec}(a) and ~\ref{result:Energy}), where melting occurs.
Initially, the recognition function and the system energy both increase simultaneously, meaning that the Pauli crystal becomes \textit{more} stable with an increase of the energy.
After reaching its maximum value, the recognition function starts to drop off as the geometric structure of the Pauli crystal becomes distorted.
This melting process is however not accompanied by a further increase in the system energy.
Instead, we see that the energy remains bounded, displaying a beating pattern around a value lower than its maximum.
Therefore, there is no evident correlation between the increase of the system energy and the melting of the Pauli crystal, at least at the experimentally relevant time scales probed by our simulations.


\subsection{Excitation spectrum}
Since there seems to be no direct correlation between an increase in the system energy and the Pauli crystal melting, we now present an alternative mechanism based on Floquet theory.

\begin{figure*}[ht]
    \centering
    \includegraphics[width=0.95\textwidth]{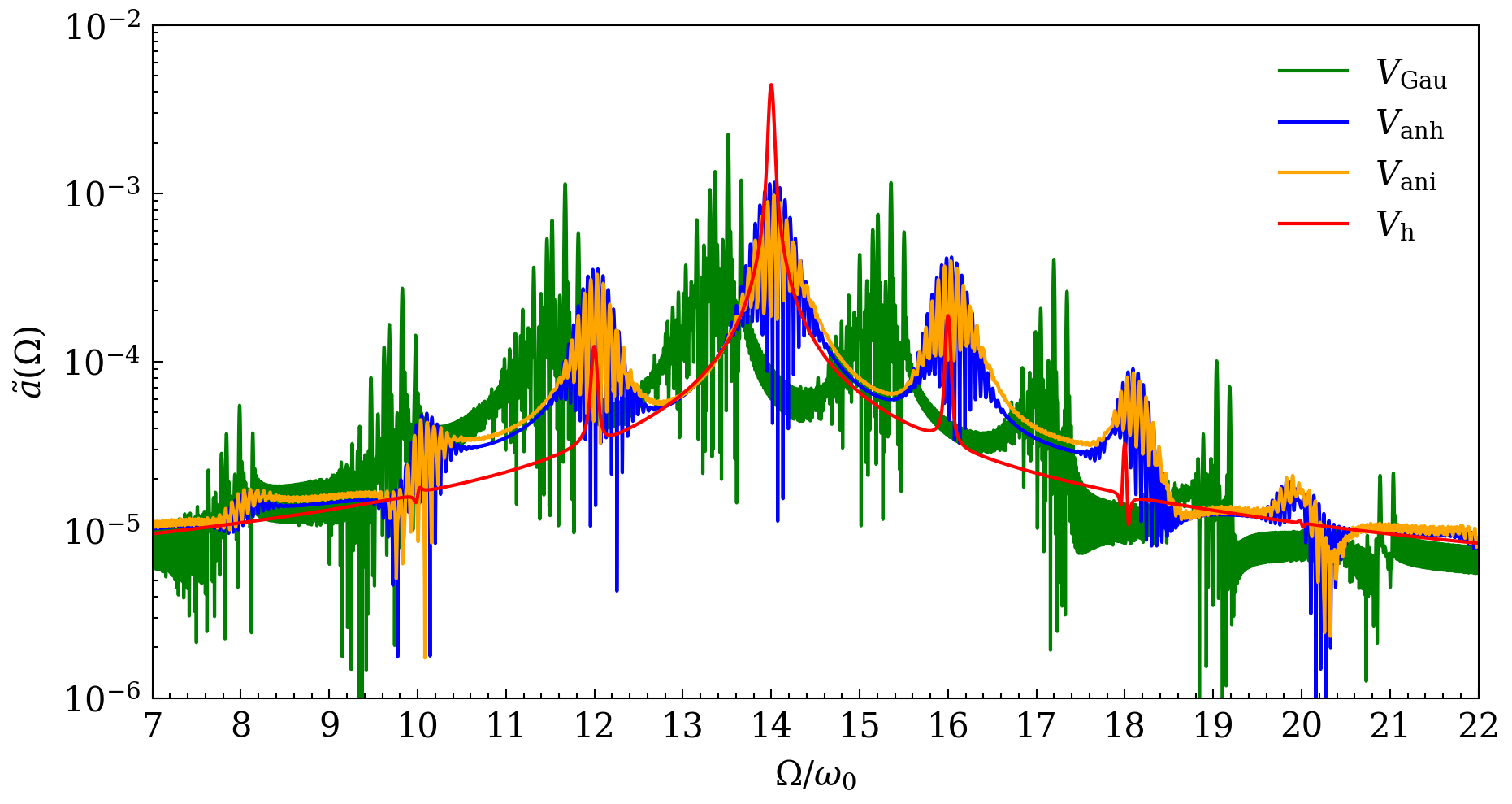}
    \caption{Excitation spectra for the six-fermion system in the four traps. The shaking amplitudes $A$ for the six-fermion system in the isotropic harmonic trap $V_\text{h}$, isotropic anharmonic trap $V_\text{anh}$, anisotropic harmonic trap $V_\text{ani}$ and Gaussian trap $V_\text{Gau}$ are $0.01,0.1,0.09$, and $0.25$, respectively.}
    \label{result:ex_spectrum}
\end{figure*}

Fig.~\ref{result:ex_spectrum} shows the excitation spectrum for the six-fermion system in the four different traps.
The composition of the quasienergy states in the spectrum of the non-melting system (isotropic harmonic trap $V_\text{h}$) and the melting systems (other traps) differs significantly.
For the isotropic harmonic trap, 
it is apparent that the largest contribution to the excitation spectrum comes from the peak with quasienergy $\alpha(j)=E_0$, \textit{i.e.} matching the ground-state energy.
The next-largest contribution comes from the two neighboring peaks at $E_0 \pm 2\omega_0$, which stem from the periodic modulation. 
Higher harmonics at $E_0 \pm  2n \omega_0$, $n \in \mathbb{Z}$ are barely visible.
Hence, the spectral density of the ground-state peak dominates the wave function in the isotropic harmonic trap.
The time-dependent system behaves very similarly to the time-independent system and the geometric structure of the Pauli crystal is preserved.
\par
In contrast, when the trap is anisotropic or anharmonic, the degenerate eigenstates of the unperturbed Hamiltonian will be split. 
Due to the energy splitting, several quasienergies with small energy differences participate in the excitation process.
The single peaks in the excitation spectrum for the six-fermion system in the isotropic harmonic trap are split into multiple peaks.
Therefore, the excitation spectrum of the melting systems is a composed of many quasienergy states with different frequencies.
The superposition of many quasienergy states sharply diminishes the contribution from the ground-state peak to the wave function.
Thereupon, the particular geometric structure observed in the ground state is gradually blurred and distorted in the modulation process.


\begin{figure*}[ht]
    \centering
    \includegraphics[width=0.95\textwidth]{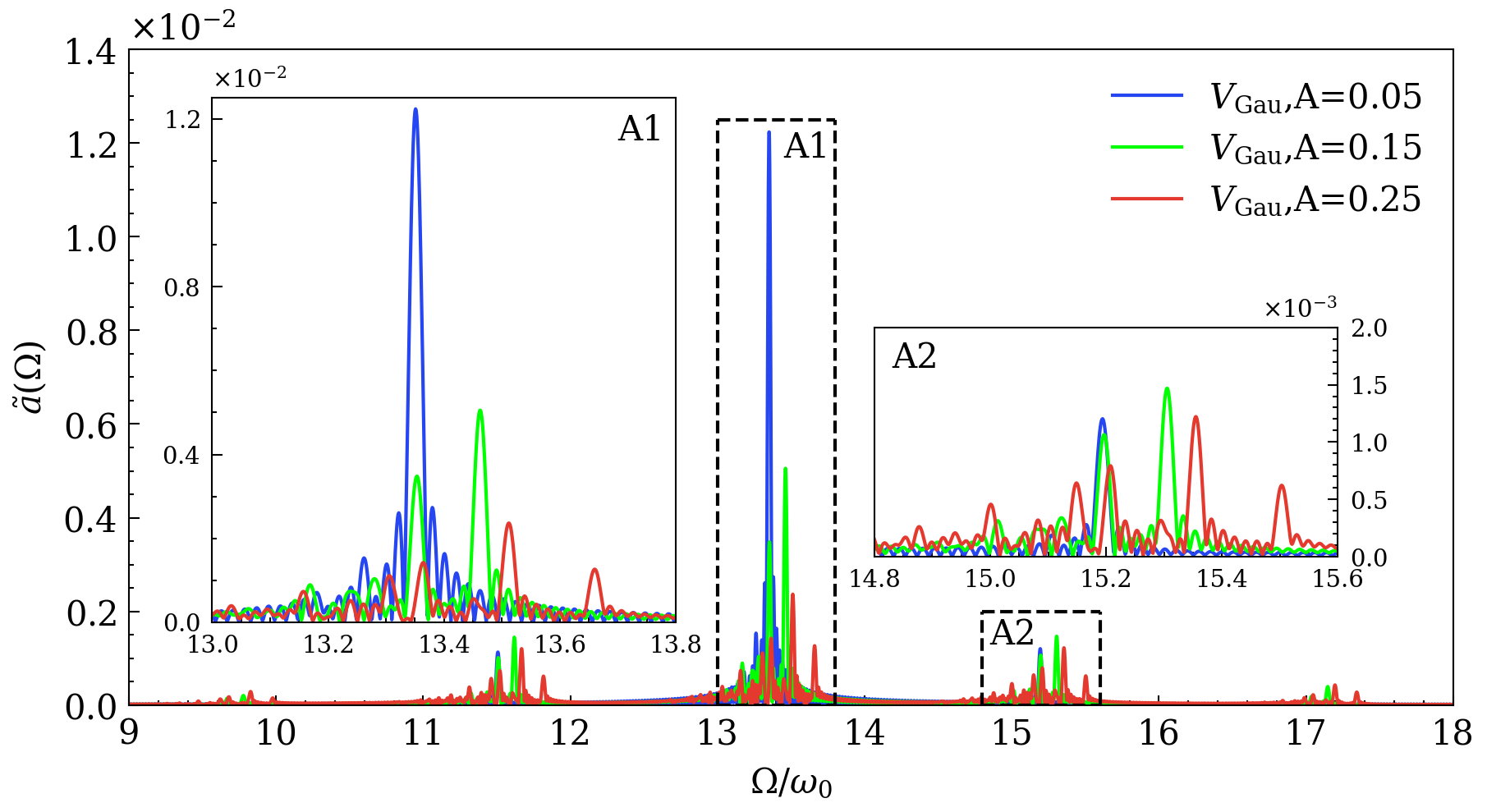}
    \caption{Excitation spectra for the six-fermion system in the Gaussian trap $V_\text{Gau}$ with different shaking amplitudes $A$.}
    \label{result:ex_spec_Gau}
\end{figure*}

We now examine the excitation spectrum for the system in the Gaussian trap $V_\text{Gau}$ in more detail.
This is shown in Fig.~\ref{result:ex_spec_Gau} for different shaking amplitudes $A$. 
In the case with a small shaking amplitude $A = 0.05$, the largest contribution to the excitation spectrum still comes from the central $E_0$-peak, while the other quasienergy states $\alpha(j)\neq E_0$ and the side peaks at $E_0 \pm  2n \omega_0$, $n \in \mathbb{Z}$ are small. 
Hence, the wave function is dominated by the ground state and the geometric structure of the Pauli correlations is still mostly visible throughout the whole modulation process. 
When the shaking amplitude increases, the ground-state contribution to the spectrum reduces while the contribution from other quasienergy states and the side peaks becomes more pronounced.
As a consequence, the wave function dynamically becomes a superposition of different quasienergy states.
The geometric structure of the Pauli crystal is destroyed completely when the shaking amplitude $A$ increases up to $0.25$.
The contribution of many quasienergy states with different frequencies drastically undermines the N-body correlation originated from the ground state.

In conclusion, the trap imperfections lead to a proliferation of multiple quasienergy states and a gradual loss of coherence in the evolution of the wave function, which manifests as the melting of Pauli crystals.

\section{\label{sec:conclusions} Conclusions}
In this work, we have studied the mechanism behind the phenomenon of Pauli crystal melting observed in recent experiments with noninteracting few-fermion systems in shaken two-dimensional optical traps.
We have simulated with MCTDH-X the many-body system dynamics in four different potentials to explore the influence of trap imperfections such as anisotropy and anharmonicity on the system dynamics.

For each setup, we have calculated the configuration density and recognition function to quantify the melting process.
By observing how the configuration density and the recognition function change with time, we classified the systems in different traps as melting or non-melting.
The melting of Pauli crystals was not observed in the perfect isotropic harmonic trap. 
For the other traps with imperfections, the melting of Pauli crystals observed experimentally was reproduced successfully. \par

To explore the reason for the melting, we have first investigated the relationship between the system energy and the melting process. 
We have shown that the melting of Pauli crystals is not a direct result of an increase in system energy. 
For instance, in a non-melting system the total energy can be much larger than in a melting system. 
We have instead found the mechanism for the melting process in the excitation spectrum calculated from the autocorrelation functions via Floquet theory.
In the isotropic, non-melting system, the many-body wave function is dominated by the ground state and the time-dependent configuration density still maintains its  correlations and geometric structure. 
In the melting systems, instead, the trap imperfections cause energy splittings which facilitate the population of new quasienergy states with small energy differences.
These are the quasienergy states involved in the excitation process and loss of coherence: the contribution of the ground state to the dynamics is diminished and the wave function dynamically becomes a superposition of different quasienergy states. 
Hence, the characteristic geometric structure of the Pauli crystal in the ground state does not dominate in the configuration density anymore and melting occurs.\par
Our study sheds light into the geometric and dynamical mechanisms that lead to Pauli crystal melting and should help devise experimental protocols that prolong the lifetime of Pauli crystals.
Reducing geometric distortions of the trap will substantially increase the stability and duration of the Pauli crystal phase.
We also remark that, since the MCTDH method is completely general, the study of melting mechanism can be extended to interacting fermionic, bosonic and multicomponent indistinguishable particles as well.
As the correlation configuration of Pauli crystals is caused by the Pauli exclusion principle, one natural question to ask is how the structure of the Pauli crystal is modified when the Pauli exclusion competes with other interactions between fermions.
Another interesting future research question would be whether structures similar to Pauli crystals could be reproduced in interacting bosonic systems, since repulsive interactions between bosonic particles might lead to a dynamical behavior similar to that of noninteracting fermionic systems.
Therefore, our study about the effect of statistics on correlation structures could give a new insight into the dynamical formation of structures such as star patterns \cite{maity2020}, Faraday waves \cite{staliunas2002}, and fireworks \cite{fu2018}.
In summary, our results show that many-body correlations offer a very rich phenomenology to explore and that dynamical excitations play a crucial role for the stability of crystalline phases of matter of quantum gases.
\begin{acknowledgments}
This work has been supported by the Austrian Science Foundation(FWF) under grants P-32033-N32.
P. M. gratefully acknowledges funding from the ESPRC Grant no. EP/P009565/1. 
We thank Rui Lin for useful comments on our manuscript.
\end{acknowledgments}
\appendix
\section{Image processing}\label{app::image}
To compare the distribution of different configurations, we can record single-shot images in such a way that only the information of the relative positions remains while other details such as the position of the center of mass and the orientation of the configuration in the space are removed. 
The ensemble of processed single-shot images is defined as the configuration density $\mathcal{S}_{N_s}$~\cite{Gajda2016}. 
The most probable configuration, namely the geometric structure of Pauli crystals, can be observed from the configuration density directly.\par
Consider the $k$-th single-shot measurement $S^{(k)}$ which collapses $N$ particles at positions $\pvec{r}_1^{(k)},\pvec{r}_2^{(k)},\cdots,\pvec{r}_N^{(k)}$. 
The positions of the $N$ particles form a configuration $D_s^{(k)}(\Vec{r})$,
\begin{equation}
    D_s^{(k)}(\Vec{r})=S^{(k)}(\pvec{r}_1^{(k)},\pvec{r}_2^{(k)},\cdots,\pvec{r}_N^{(k)}).
\end{equation}
The function $D_s^{(k)}(\Vec{r})$ is only non-zero at positions $\pvec{r}_1^{(k)},\pvec{r}_2^{(k)},\cdots,\pvec{r}_N^{(k)}$ in the measurement $k$. 
The average of $N_s$ repeated single-shot measurements forms a histogram $\mathcal{D}_{N_s}$ of atomic position in a  two-dimensional plane:
\begin{equation}
    \mathcal{D}_{N_s}(\Vec{r})=\frac{1}{N_s}\sum_{k=1}^{N_s}D_s^{(k)}(\Vec{r}).
\end{equation}
It is intuitive that the direct summation of single-shot images yields the one-body density in the limit of infinite repeated measurements,
\begin{equation}
    \lim_{N_s\to \infty} \mathcal{D}_{N_s}(\Vec{r})=\rho(\Vec{r}),
\end{equation}
where $\rho(\Vec{r})$ is the one-body density
\begin{equation}
    \rho(\Vec{r})=\frac{\bra{\Psi}\hat{\Psi}^\dagger(\Vec{r})\hat{\Psi}(\Vec{r})\ket{\Psi}}{N}.\label{App:one-body}
\end{equation}
This quantity cannot provide information about correlations between particles.

\par
In order to reveal the correlation configuration, we do additional image processing.
The first step is to shift the center of mass of the configuration to the origin of the coordinate system:
\begin{equation}
    \pvec{r}_i^{\prime (k)}=\pvec{r}_i^{(k)}-\pvec{r}^{(k)}_{CM},
\end{equation}
where $\pvec{r}^{(k)}_{CM}=\frac{1}{N}\sum_{i=1}^N \pvec{r}_i^{(k)}$. 
On the account of the randomness in the orientation, the single-shot images require applying a rotation with specified angles depending on different configurations. 
Thus, we can set a fixed configuration $\{\vec{r}_0\}_N$ as the reference configuration. 
The reference configuration can be chosen arbitrarily but is the same for every single-shot image. 
For convenience, we use polar coordinates instead of Cartesian ones, 
$\vec{r}_{0,i}\to (r_{0,i},\phi_{0,i})$.
The distance between the measured configuration $\{\pvec{r}^\prime\}_N$ and the given reference configuration $\{\vec{r}_0\}_N$ is defined as the angular difference
\begin{equation}
    d(\{\pvec{r}^\prime\}_N, \{\vec{r}_0\}_N)=\sum_{i=1}^N(\phi_i^\prime-\phi_{0,i})^2.
\end{equation}
Then, we rotate the $k$-th measured configuration after the translation to $\{\pvec{r}^\prime\}_N^{(k)}$ by a specified angle $\alpha_k$ to minimize the distance between it and the given reference configuration $\{\vec{r}_0\}_N$:
\begin{equation}
    \alpha_k = \mathrm{argmin} \sum_{i=1}^N(\phi_i^\prime-\alpha_k-\phi_{0,i})^2.
\end{equation}
By taking the derivative of the distance, it is straightforward to get the value of $\alpha_k$:
\begin{equation}
    \alpha_k=\frac{1}{N}\sum_{i=1}^N (\phi_i^\prime-\phi_{0,i}).
\end{equation}
Since the angles $\phi_{0,i}$ of the reference configuration $\{\vec{r}_0\}_N$ are constant for all single-shot images, they  only influence the orientation of the pattern of the configuration density and not its geometric structure. 
Hence, we can set $\phi_{0,i}=0$ and the rotation angle $\alpha_k$ is the average of all angles $\phi_{k,i}^\prime$:
\begin{equation}
    \alpha_k=\frac{1}{N}\sum_{i=1}^N\phi_{k,i}^\prime.
\end{equation}
The maximal rotation angle $\alpha_k$ must be limited to $2\pi/l$ when the system obeys a $l$-fold symmetry. 
And the new coordinates after the rotation in the $k$-th single-shot image are:
\begin{equation}
    \Tilde{\vec{r}}_{k,i}=(r_{k,i}^\prime, \phi_{k,i}^\prime-\alpha_k)\quad \alpha_k\in \left[0,2\pi/l \right].
\end{equation}
The configuration density can finally be obtained by adding up all processed single-shot images:
\begin{equation}
    \mathcal{S}_{N_s}(\vec{r})=\frac{1}{N_s}\sum_{k=1}^{N_s} D_s^{(k)}(\Tilde{\vec{r}}).
\end{equation}

\section{Parameter setting}\label{app::units}
\begin{table}[h]
    \centering
    \begin{tabular}{|c|c|c|c|}
    \hline
    Trap & Parameter & Range & Step size\\
    \hline
    $V_h$& $A$ & From 0.001 to 0.01 & 0.001 \\
    \hline
    \multirow[c]{2}{*}{$V_{ani}$} & $A(\gamma=0.9)$ & From  0.01 to 0.09 & 0.01\\
    \hhline{|~|---|}
    & $\gamma(A=0.01)$ & From  0.90 to 0.99 & 0.01\\
    \hline
    \multirow[c]{2}{*}{$V_{anh}$} & $A(B=0.001)$ & From 0.01 to 0.1 & 0.01\\
    \hhline{|~|---|}
    & $B(A=0.1)$ & From 0.001 to 0.01 & 0.001\\
    \hline
    $V_{Gau}$  & $A(B^\prime=20, \gamma^\prime=0.99)$ & From 0.01 to 0.25 & 0.01\\
    \hline
    \end{tabular}
    \caption{Parameter space in simulations. The parameter A is the shaking amplitude given in the Eq.~\eqref{modulation}. $\gamma$ and $B$ are parameters to quantify the anisotropy and anharmonicity of the trap, respectively. In the Gaussian trap, the anisotropy $\gamma'$ and anharmonicity parameter $B'$ are experimental values.}
    \label{tab:para}
\end{table}
Table~\ref{tab:para} shows the parameter space that we covered in our simulations. We simulated the modulation process in four different traps with different sets of parameters. 
The energy input into the system is governed by the magnitude of the shaking amplitude $A$. 
The value of $A$ ranges from 0.01 to 0.1 in the experiment. 
We strive to use the same shaking parameters as the experiment in the simulation. 
However, simulations of the isotropic harmonic trap with shaking amplitudes $A>0.01$ are not feasible due to a spatial extent problem. 
For those values, since the system in the isotropic harmonic trap is in resonance with the modulation frequency, the one-body density of the system spreads outwards rapidly and shoots out of the spatial domain that we consider.
To solve this problem, the value range of the shaking amplitude in the isotropic harmonic trap is adjusted to the interval $[0.001, 0.01]$ with step size 0.001. Similarly, the simulation in the anisotropic harmonic trap with parameters $\gamma=0.9$ and $A = 0.1$ can not be implemented as a consequence of the same issue.\par

In our computations we have adopted dimensionless units.
An initial trap frequency of $\omega_0=1$ in our computations corresponds to $\omega_0=2\pi\times985(3)$Hz in the experiment. 
The modulation time $t=309$ in the computations  corresponds to $50ms$ in the experiment. 
The overall parameter settings in the computations and all simulation results can be accessed through the \href{https://gitlab.com/Jiabing/pauli-crystals.git}{online data repository}.

\section{Floquet Theory}\label{app::floquet}
According to the Floquet theorem \cite{Floquet1883}, the solution to the Schr\"odinger equation can be written in the following form:
\begin{equation}
    \Psi_\alpha(\vec{r},t)=e^{-i\epsilon_\alpha t}\varphi_\alpha(\vec{r},t),
\end{equation}
where $\varphi_\alpha(\vec{r},t+\tau)=\varphi_\alpha(\vec{r},t)$. 
The wave function $\Psi_\alpha(\vec{r},t)$ is called the quasienergy state (QES) and $\epsilon_\alpha$ is a real parameter called the Floquet characteristic exponent or the quasienergy. 
If we insert the solution $\Psi_{\alpha}(\vec{r},t)$ into the Schr\"odinger equation, we find:
\begin{equation}
    \hat{\mathcal{H}}(\vec{r},t)\varphi_\alpha(\vec{r},t)=\epsilon_\alpha\varphi_\alpha(\vec{r},t).
\end{equation}
where $\hat{\mathcal{H}}\equiv\hat{H}_\tau(\vec{r},t)-i\partial/\partial_t$.
On the account of the periodicity of $\varphi_\alpha(\vec{r},t)$, the quasienergy state (QES) function $\Psi_\alpha(\vec{r},t)$ can be then expanded in a Fourier series \cite{Chu2004},
\begin{equation}
    \Psi_\alpha(\vec{r},t)=e^{-i\epsilon_\alpha t}\sum_{n \in \mathbb{Z}} \varphi_{\alpha }^n(\vec{r})e^{-in\omega t},
\end{equation}
where
\begin{equation}
    \varphi_{\alpha }^n(\vec{r})=\int \mathrm{d} t \varphi_\alpha(\vec{r},t)e^{in\omega t}.
\end{equation}
For the generalized Hamiltonian $\hat{\mathcal{H}}(\vec{r},t)$, we can introduce the composite Hilbert space $\mathcal{L}=\mathcal{R}\otimes\mathcal{T}$ \cite{Shirley1965,Sambe1973} which is spanned by any set of square-integrable functions in the spatial space $\mathcal{R}$ and the complete orthonormal set of functions $\{\exp(in\omega t)\}$, $n \in \mathbb{Z}$ in the temporal space $\mathcal{T}$. 
The notation $|\Psi\rangle\rangle$ represents a vector in the composite space $\mathcal{L}$ and the inner product is defined as
\begin{equation}
    \cbraket{\Psi_\alpha(\vec{r},t)}{\Psi_{\alpha^\prime}(\vec{r},t)}=\frac{1}{\tau}\int_0^\tau \bra{\Psi_\alpha(\vec{r},t)}\ket{\Psi_{\alpha^\prime}(\vec{r},t)}.
\end{equation}
And the eigenvectors of $\mathcal{H}(\vec{r},t)$ form a complete set in the composite space $\mathcal{L}$:
\begin{equation}
    \sum_\alpha|\Psi_\alpha\rangle\rangle\langle\langle\Psi_\alpha|=I.
\end{equation}
Hence, the wave function of the time-dependent system with the time-periodic Hamiltonian $\hat{H}_\tau(\vec{r},t)$ can be formally expanded as a superposition of QES:
\begin{align}
    \Psi(\vec{r},t)&=\sum_j A_j\Psi_{\alpha(j)}(\vec{r},t)\label{eq26}\\
    &=\sum_j\sum_{n=-\infty}^\infty A_j \varphi_{\alpha(j)}^n(\vec{r})e^{-i(\epsilon_{\alpha(j)}+n\omega)t}.\label{App:floquet-expansion}
\end{align}

\nocite{*}

\bibliography{ref}

\end{document}